# Bit Copying – The Ultimate Computational Simplicity

Oleg Mazonka, July 2009
mazonka@gmail.com

*Abstract.* A computational abstract machine based on two operations: referencing and bit copying is presented. These operations are sufficient for carrying out any computation. They can be used as the primitives for a Turing-complete programming language. The interesting point is that the computation can be done without logic operations such as *AND* or *OR*. The compiler and emulator of this language with sample programs are available on the Internet.

## *Introduction*

In a quest to build an imperative language with the smallest possible number of instructions several One Instruction Set Computer (OISC) languages have been invented. One example, the ultimate RISC architecture [1], utilizes a single instruction, copy memory to memory. The complex behaviour of such machine is achieved by mapping the machine registers on memory cells. For example, a memory cell with the address zero is the instruction pointer, so copying to this address effectively realizes unconditional jump. Arithmetic operations are also achieved by using special registers in memory performing more complex operations on hardware level.

Another example, Subleq [2], does not have memory mapped registers. Its computational power based on program self-modification and its sufficiently complex instruction. The abstract machine is defined as a process working on an infinite array of memory cells with each instruction having three operands. The processor reads from the memory three operands, subsequent cells *A B C*, subtracts the value of the cell addressed by *A* from the value of the cell addressed by *B* and stores the result in the same cell addressed by *B*. If the result is less than or equal to zero, the execution jumps to the address *C* and the processor reads the next instruction from there, otherwise next 3 operands are read from the memory. This language is proven to be Turing-complete. There are a few variations of this language, which are similar in principle. A compiler from a simple C-like language has been written, which compiles a program into Subleq processor code [3]. Attempts to reduce the complexity of the atomic operation had been undertaken. For example, Rojas [5] proves that conditional branching is not necessary for universal computation given the ability of code self-modification.

Although OISC languages have just one instruction, the instruction does a number of manipulations or computations under the hood. Hence there is a question: which language has the simplest instruction and is it possible to make a language with a simpler instruction?

Another interesting question relates to logic operations. It is commonly known that any classical computations are usually done using bit logic operations: *AND*, *OR*, *XOR*, and *NOT*. These operations are neither a complete set nor a minimal set required for computation. *OR* and *XOR* can easily be expressed via `AND` and `NOT` and vice versa. However it is commonly assumed that one needs at least `AND` or `OR` like operations to make real computations. It is not possible to combine `OR` and `XOR` operations to erase a single bit. Therefore they are unable to produce classical computations. Hence a question arises: is it possible to make programmed computation without using logical operations like `AND` and `OR`?

## *1. Referencing as Computational Operation*

Surprisingly, `OR` and `XOR` reversible operations can produce irreversible results if they are used in combination with referencing. In the following table:

| 000 | 001 | 010 | 011 | 100 | 101 | 110 | 111 |
|-----|-----|-----|-----|-----|-----|-----|-----|
| 100 | 011 | 011 | 111 | 110 | 100 | 010 | 101 |

the first row has initial three bits. The second row has the same bits with one inverted (`NOT` operation applied). The one inverted is the one referenced by all three, as the index of the bit equal to binary representation taken modulo 3. As one can see two initial states (001 and 010) produce the same final result (011), which makes this entire operation irreversible.

A machine, similar to register machines described in Stephen Wolfram's NKS [4], can be realized using continuous process of bit inversion on the same set of bits. For example, 3 bit machine produce a sequence (000) (100) (110) (010) (011) (111) (101) (100) ...

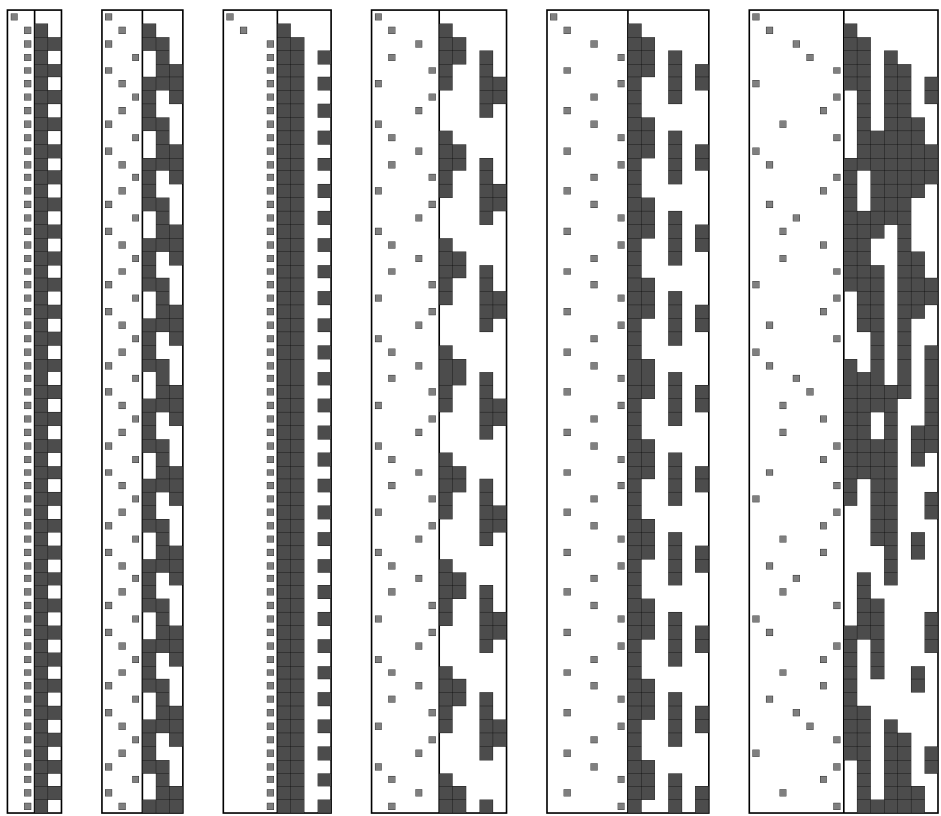

The figure above shows Wolfram diagrams for 2, 3, 4, 5, 6, and 7 bit machines. On the right side of each diagram bits represented as dark (for 1) and white (for 0) squares. On the left side a small square shows the interpreted value of the bits – the address of the next bit to be inverted. The address is calculated as the binary representation of some integer taken modulo number of bits.

Taking binary value and modulo may seem like complex operations. However there is a simple model simulating such behaviour. Allow bits positioned in a circle and an arrow pointing to any bit. Each bit is assigned a rule how to rotate the arrow if its value is 1. In one step the arrow rotates according to the rules and the values of all bits, and at the end of the step the bit pointed by the arrow is inverted.

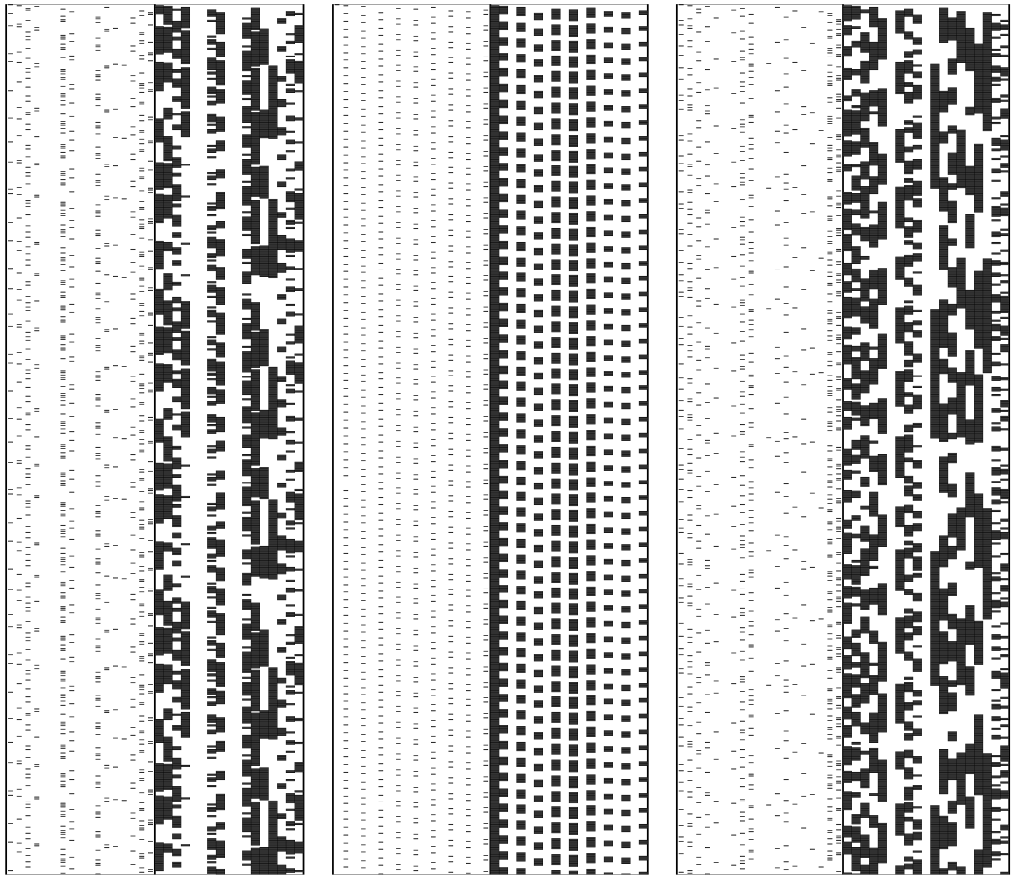

This Figure above shows the diagrams for 17, 18, and 19 bit machines similar to the previous figure. The picture is scaled 1 to 10 in the vertical direction. 1000 steps are shown.

The table below shows the size of the loop (the size of the pattern on the diagram) the machine enters eventually when started with the initial zero values of all bits.

| Bits | Loop size | Bits | Loop size | Bits | Loop size |
|---|---|---|---|---|---|
| 3 | 6 | 13 | 66 | 23 | 18812 |
| 4 | 2 | 14 | 50 | 24 | 6 |
| 5 | 8 | 15 | 162 | 25 | 48000 |
| 6 | 6 | 16 | 2 | 26 | 544 |
| 7 | 50 | 17 | 346 | 27 | 54 |
| 8 | 2 | 18 | 18 | 28 | 62 |
| 9 | 18 | 19 | 1700 | 29 | 128116 |

| 10 | 10  | 20 | 10  | 30 | 30     |
|----|-----|----|-----|----|--------|
| 11 | 112 | 21 | 12  | 31 | 635908 |
| 12 | 6   | 22 | 118 | 32 | 2      |

Bigger values of the loop size correspond to more complex behaviour of the model.

In another example

| 0000 | 0001 | 0010 | 0011 | 0100 | 0101 | 0110 |
|------|------|------|------|------|------|------|
| 0000 | 0001 | 1010 | 1011 | 1100 | 0001 | ...  |

the second row has one bit modified by the formula

$$(\ p[A]\ XOR\ p[B]\ ) \rightarrow p[A],$$

where **A** is the value of the first two bits, **B** is the value of the last two bits, and **p[ ]** is the bit taken by index.

The operations applied to bits are reversible, but referencing makes the whole process irreversible. Even if it feels like this process can do calculation, it is still difficult to bring it into play to make a framework able to do desirable computations.

## *2. Bit Copying Language*

It turns out that by taking the converse approach – that is to combine a bit copying operation, which is irreversible, with referencing – the programmable computation is possible at the bit level.

A bit copying instruction always erases one bit. On first inspection it would seem that the entire amount of information in the system is forever reducing until no changes are made. However this is not the case. For example, in the process: *forever do (a→b, c→a, b→c)*, where *a*, *b*, and *c* are bits, and arrow means copy, the bits will circle in the loop forever. This would be interpreted as a steady but not static state. To do something more interesting lets add a meaningful representation to each collection of bits by associating an address with each.

Let us define the imperative language, in which the abstract machine operates on an array of memory bits addressed from zero. Bits are grouped into *word*s of a particular size, memory cells. For example, 8-bit words:

| Memory  | 01010101 | 00001111 | 10101010 | 11001100 | 00110011 | ... |
|---------|----------|----------|----------|----------|----------|-----|
| Address | 0        | 8        | 16       | 24       | 32       | 40  |

Each instruction consists of three operands: **A B C**, where each operand is one word. The instruction copies the bit addressed by **A** into the bit addressed by **B**, then jumps the process control to the address **C**. The operand **C** is read after the bit copying is done. This allows the instruction to be self-modifiable (even though just one bit can be modified at a single point in time).

Since each word represents a bit address in the memory, the exact way of address interpretation can be left undefined without sacrificing the concept. In this implementation, however, it is assumed little-endian binary bit representation.

Another undefined up to implementation feature is how to consider grouping memory bits into words: physical – all words are aligned on memory cell size address, or logical – words are grouped counting from the current address. In the second case the operand *C* is allowed to specify any address, not only multiple to the memory cell size.

This sole instruction does not do any more than just copying a bit from one place to another, and yet this simple single instruction is enough to make the language able to execute pre-programmed sequence of operations. The abstract machine obviously does something more than copying bits: it references bits and transfers the process control to the next address of execution. However this work may not need logic operations *AND* or *OR*, and is done outside of the execution model, which means it can be emulated by the same bit copying process.

## *3. Assembly*

To simplify the presentation of the language instruction let us use the following assembly notation. Each word is denoted as `L:V'x`, where `L` is an optional label serving for addressing this memory cell, `V` is the value of this memory, `x` is the optional bit offset within the word (memory cell). Each instruction is written on a separate line. For example:

```
A'0 B'1 A
A:18 B:7 0
```

There are two instructions above. The first instruction copies the lowest bit of the cell *A* (whose value is 18) into the second bit of the cell *B* (whose value is 7), then jumps to the instruction addressed by *A*, which is the next instruction. After the first instruction is executed the value of the cell *B* is changed to 5. For example, in 8-bit word memory these two instructions are

```
24 33 24 18 7 0.
```

If the bit offset is omitted it is considered to be 0. So, *A* is the same as *A'*0.

If the operand *C* is absent it is assumed to have the value of the next cell address, i. e.

```
A B
```

is the same as

```
A B C
C: ...
```

which is the same as

```
A B ?
```

Question mark (**?**) is the address of the next memory cell, i.e. the address of the first bit of the following word. Let's denote (**n?**) as a multiple to cell size counted from the current position. So, (`0?`) means the address of this position; (`1?`) is the same as (`?`) and is the next cell; (`2?`) is one after the next cell; and (`-2?`) is one before the previous cell. For example, the instruction

```
A B -2?
```

is the same as

```
C:A B C
```

and is an infinite loop (assuming that it does not modify `C`), as the bit referenced by `A` is copied to the bit referenced by `B` and the control is transferred to the address of the cell `C:A`, which is the beginning of the original instruction. Remember that `A` is the value and `C` is the address of `C:A`.

Given an assembly text one needs an environment to run a program, which is written in this language. Therefore two steps are necessary: 1) Compile the text into binary code as an array of bits to form instructions; 2) Execute the binary coded instructions on the abstract machine. A program called an assembler does the first step, and an emulator can do the second.

## *4. Macro Commands*

To make a program description shorter and more readable let us define a macro substitution mechanism as in the following example:

```
.copy A B
...
.def copy X Y
X'0 Y'0
X'1 Y'1
...
X'w Y'w
.end
```

The first line is the macro command which is substituted by the body of macro definition starting with ".`def`" and ending with ".`end`". The name after ".`def`" becomes the name of the macro and all subsequent names are formal arguments to the macro. After macro substitution the code becomes:

```
A'0 B'0
A'1 B'1
...
A'w B'w
```

Here $w$ is the index of the highest bit. It is equal to the size of memory cell minus one. Lets denote the word size as $W$, $W=w+1$.

Two other useful macro definitions are shift and roll. `shiftL` shifts bits in the memory cell by 1 from lower to higher, and the lowest bit is set to 0. It is the same as arithmetic multiplication by 2, or the C programming language operation "`<<=1`".

```
.def shiftL X : ZERO
X'(w-1) X'w
X'(w-2) X'(w-1)
...
X'1 X'2
X'0 X'1
```

```
ZERO X
.end
```

`ZERO` is defined at some place as `ZERO:0`. It appears after the colon at the end of macro definition argument list to signify that this name is defined outside the macro definition. This is necessary because the assembler tries to resolve all names within the body of macro definition or to tie them to the formal arguments. Note, that the last instruction copies the lower bit of the `ZERO` memory cell to the lower bit of *x*.

`shiftR` is the same as `shiftL` but works in the opposite direction. It corresponds to integer division by 2, or the C shift operator "`>>=1`".

Roll macros are similar to shift macros with the exception that they copy the end bit back to the front. They can be defined in terms of the shift macro definitions:

```
.def rollR X : TMP
X TMP
.shiftR X
TMP X'w
.end
.def rollL X : TMP
X'w TMP
.shiftL X
TMP X
.end
```

The `TMP` memory cell is a placeholder and is defined in an external library.

The macros copy, shift, and roll are useful, but lack the logic to be able to perform useful calculations.

## 5. Conditional jump

Consider the following code:

```
.def jump01 A b
A'b 2?'k
0 J'0
A'b 2?'k
1 J'1
A'b 2?'k
2 J'2
...
A'b 2?'k
(w-2) J'(w-2)
A'b 2?'k
(w-1) J'(w-1)
A'b 2?'k
w J'w J:0
.end
```

The offset $k$ is defined such that $2^k=W$. Since a word is an address of a bit in the memory, there are $k$ bits corresponding to the offset within a word. The rest of the word's bits specify the address of a memory cell. For example, for a 32-bit word $k$ is 5 because modifications in the sixth bit and higher change the address of the memory cell, but not the offset inside the memory cell. Note, that when writing to the offset $k$ updates the sixth bit if $k$=5.

The first line moves the bit **b** of the memory cell **A** to the *k*-th bit of the first operand of the next instruction. After this is done the first operand of the next instruction is zero or equal to **W**. The next instruction copies the value of the first bit of the cell addressed either zero or **W** to the cell labelled **J** – which is the last memory cell in this list of instructions, and which is the address where the process control will go after the last instruction is executed. The subsequent lines [(A'b 2?'k)(1 J'1)] copy the second bit to the **J** cell, and so on.

When the last bit is copied, the cell **J** holds the same value as the cell with address 0 (the first word in the memory) or the cell with address **W** (the second word in the memory) depending on whether **A'b** was 0 or 1.

By marking the first two memory cells in the program as special, in the sense that they can be used only for this operation, it is possible to write a generic test for a particular bit.

```
        Z0:0 Z1:0
        .def test A b B0 B1 : Z0 Z1
        .copy L0 Z0
        .copy L1 Z1
        .jump01 A b
        L0:B0 L1:B1
        .end
```

This code defines a macro that tests the bit **b** of memory cell **A** and jumps to either address **B0** or **B1** if the bit is 0 or 1 correspondingly.

Testing a bit is a core requirement of all the higher-level computation described below. In most of these cases testing a bit involves checking the lowest or highest bit in the word:

```
        .def testL A B0 B1
        .test A 0 B0 B1
        .end
        .def testH A B0 B1
        .test A w B0 B1
        .end
```

## 6. Arithmetic

One of the most basic operations, which will be required for definitions of other more complex constructions, is the increment operation. To increment a memory cell **A** one can combine the roll, shift, and test macros in the following way:

```
        .copy ONE ctr

begin:  .testL A test0 test1

test0:  ONE A rollback
test1:  ZERO A
        .testH ctr next rollback

next:   .shiftL ctr
        .rollR A
        0 0 begin

rollback: .testL ctr roll End
```

```
roll:   .shiftR ctr
        .rollL A
        0 0 rollback

        End:0 0
        ...
        ctr:0 0
```

The first line initialises the counter `ctr` to 1. The lowest bit of the operand **A** is swapped, and the operand and counter are rolled until either operand bit is zero or the counter bit 1 reaches the highest bit position, which means that all *w* bits of the operand were processed. After this the operand can be rolled back to the original bit position.

In the code above `ZERO` and `ONE` are defined as `ZERO:0` and `ONE:1`. The instruction `0 0 label` is used as an unconditional jump to address `label`. It copies the first bit in the memory to itself, and does not change its value.

Addition can be defined in a similar way with the exception that there are four operands. These are in order: first number, second number, result, and the adder (for passing over an extra bit). The lowest bits of the adder and first and second numbers are added giving two bits, one of which goes to the lowest bit of the result, and the other goes to the second bit of the adder. Then all four operands are rolled and the process continues. The exact code for addition command `add` is given in Appendix B1.

Subtraction can simply be defined as

```
.def sub X Y Z
.inv Y
.inc Y
.add X Y Z
.end
```

where `inc` is increment, `add` is addition (*Z=X+Y*) described above, and `inv` is the inversion operation, which simply inverts all bits in a memory cell. The inversion operation is simpler than increment. See Appendix B2 for the `inv` code definition.

## 7. Process Control and Pointers

To test for particular values of variables one can use the following definitions:

```
.def ifeq X Y yes no
.sub X Y Z
.ifzero Z yes no
Z:0 0
.end

.def ifzero Z yes no
.testH Z cont no
cont: .copy Z A
.inv A
.inc A
.testH A yes no
A:0 0
.end

.def iflt A B yes no
```

```
        .sub A B Z
        .testH Z no yes
        Z:0 0
        .end
```

The first macro `ifeq` checks if both arguments equal to each other by testing the result of the subtraction. The second macro `ifzero` tests whether its argument is equal to zero. This is performed in the following way. Test the highest bit (negative value), if this bit is zero – negate the argument (in a simple binary signed representation inversion and increment produce the same result as negation) and re-test the highest bit. Note, that the argument is copied before it is negated because it should not be changed. The third macro `iflt` tests if the first argument is less than the second.

To write the classical "Hello, World!" program, by iterating a pointer over an array of cells, one needs to define the tricky operation of pointer dereferencing. Consider the following program:

```
        Z0:0 Z1:0

start:  .deref p X
        .testH X print -1
print:  .out X
        .add p W p
        0 0 start

        p:H X:0
        H:72 101 108
        108 111 44
        32 87 111
        114 108 100
        33 10 -1
```

The label *H* is the address of a string holding the ASCII code for "Hello, World!" followed by the end-of-line sentinel. *p* is a pointer – a cell initialized with the address of the string.

The first instruction does not do anything since it copies the bit addressed 0 to itself. It is necessary because the conditional jump (which uses the first two words of the memory) is part of other macro commands. The next command dereferences *p* by copying the value of the cell, whose address is stored in *p*, into the cell *X* (this operation is discussed below). Check if *X* is negative. If it is go to the address of (**-1**), otherwise continue execution to the next line. The address (**-1**) is special in the sense that it is assumed that the program halts if the control is passed to the address (**-1**). In fact, this is similar to how halt is defined in other one instruction set languages, for example, Subleq [2]. The next line prints the ASCII character in cell *X*. [The specific implementation of printing will be discussed later in the input/output section.] If *X* was not negative, the pointer *p* has not reached the end of array and still points to a valid array element. The pointer is incremented by the size of the memory cell and this process is continued until the halting instruction is executed.

It is possible to copy a memory cell referenced by another memory cell by setting up an iterative instruction with the source and target addresses and repeat this instruction for *W* times incrementing the addresses each time, so the whole word is copied. For example, like in the following code:

```
        .copy ONE ctr
        .copy P A
        .copy L B

begin:  A:0 B:0
        .testH ctr next End
```

```
next:    .shiftL ctr
         .inc A
         .inc B
         0 0 begin

         End: ...
         L:X ctr:0
```

This block of code does the same as C programming language statement **X=*P**. The counter is prepared as in the previous examples. The pointer value is copied to the first operand of the iterative instruction (`A:0 B:0`), then the address of the result cell is copied into the second operand of the iterative instruction. Now the iterative instruction is executed *W* number of times with each execution incrementing the addresses – values of the operands.

This approach can be used for copying a value into a memory cell pointed by another pointer. It is just a matter of swapping the **A** and **B** operands in the iterative instruction.

## *8. More Arithmetic*

Multiplication is quite simple once shift and addition are implemented[1]:

```
         .copy ZERO Z

begin:   .ifzero X End L1
L1:      .testL X next L2
L2:      .add Z Y Z
next:    .shiftR X
         .shiftL Y
         0 0 begin

         End:0 0
```

This code shifts the first multiplier to the left and the second multiplier to the right at the same time accumulating the result by adding the second multiplier if the lowest bit of the first multiplier is 1. This algorithm is expressed in a simple formula:

$$X \times Y = \begin{cases} X/2 \times 2Y, & \text{if } X \text{ even}; \\ (X-1)/2 \times 2Y + Y, & \text{if } X \text{ odd}. \end{cases}$$

Division is slightly more complex. Given two numbers *X* and *Y*, increase *Y* by 2 until the next increase gives *Y* greater then *X*. At the same time as increasing *Y*, increase a variable *Z* by 2, which is initialised to 1. Now *Z* holds the part of the result of division - the rest is to be calculated further using *X-Y* and *Y*, which is done iteratively accumulating all *Z*'s. At the last step when *X<Y*, *X* is the remainder. Code of the division is presented in Appendix B3.

The division operation is imperative for printing numbers as decimal strings. The algorithm implementing this divides the value by 10 and stores the remainders into an array. When the value becomes 0, it iterates backwards over the array printing numbers in ASCII code.

```
         .testH X begin negate

negate:  .inv X
```

---

[1] The algorithm does not properly handle negative values. This is sacrificed for the sake of simplicity.

```
                .inc X
                .out minus
begin:   .div X ten X Z
                .toref Z p
                .add p W p
                .ifzero X print begin

print:   .sub p W p
                .deref p Z
                .add Z d0 Z
                .out Z
                .ifeq p q End print

                End:0 0
                ...
                Z:0 d0:48 ten:10
                p:A q:A minus:45
                A:0 0 0
                ...
```

The first section, labelled `negate`, checks whether the argument is less than 0. If so, then the argument is negated and the minus sign is printed. The second section repeatedly divides the argument and stores the results into the array **A** by a dereferencing operation through the pointer **p**. The command `div` divides **X** by 10, stores the result back to **X** and the remainder to **Z**. The following command `toref` writes the value of **Z** into the cell pointed by **p**. This process continues until **X** is zero. In the next section marked by the label `print` the pointer **p** runs back until it is equal to **q**, which is initialised to **A**, which is the beginning of the array. The command `deref` copies the value from the array to **Z**. Then the ASCII code (48) for character 0 is added and the byte is printed. [It is assumed that the memory cell is not less than 8-bit byte.]

## 9. Input and Output

At this point, it is possible to write a program that can add, subtract, multiply, divide, iterate, dereference, and jump. To produce an output or receive an input, one has to define *what* is the output and input. This is called the pragmatics of the language or the environment of the abstract machine, which implements the language. Any definition of input to or output from the abstract machine will be a burden of the environment, or in our case the emulator of the language (or processor if implemented as hardware). Since the program can copy only bits, it is natural to define a stream of bits as bits copied to or from a particular address. One special address (`-1`) has already been introduced as the halt address – a program halts if the process control is passed to the address (`-1`). One can use the same address without ambiguity:

```
                .def out H
                H'0 -1
                H'1 -1
                H'2 -1
                H'3 -1
                H'4 -1
                H'5 -1
                H'6 -1
                H'7 -1
                .end

                .def in H
                -1 H'0
                -1 H'1
                -1 H'2
```

```
        -1 H'3
        -1 H'4
        -1 H'5
        -1 H'6
        -1 H'7
        .end
```

Note, that only the lower eight bits are copied to and from the word. This is for practical reasons. With this definition it is possible to write a word size independent assembly code, which inputs and outputs characters as 8-bit symbols.

The emulator keeps buffers of up to eight bits. When the program outputs a bit, it is placed into the buffer. When the buffer is full, a character in ASCII code is flushed to the standard emulator's output from the buffer. When the program copies a bit from the input, it is removed from the input buffer; and if the buffer is empty a character is read and its bits are placed into the buffer.

Below is a program, which prints the first twelve factorials.

```
        Z0:0 Z1:0

start:  .prn X
        .mul X Y Y
        .out ex
        .out eq
        .prn Y
        .out eol
        .inc X

        .ifeq X TH -1 start

        X:1 Y:1 ex:33
        eol:10 eq:61 TH:13
```

The macro `prn` is a printing command described in the previous section. The output of the program is

```
1!=1
2!=2
3!=6
4!=24
5!=120
6!=720
7!=5040
8!=40320
9!=362880
10!=3628800
11!=39916800
12!=479001600
```

This program runs sufficiently quickly on a modern computer with the current implementation of the assembler, emulator, and a collection of macro-defined commands. The word size is 32 bits and the size of the program (after assembling) is about 10,000 instructions.

## *10. Functions and Library*

It is handy to put all macro definitions into one file – a library, and use it with any program. For this, a third keyword command is defined (the other two are `def` and `end`):

```
.include library_file_name
```

Any program using the library is required to include it and start with the line (`Z0:0 Z1:0`). For example,

```
Z0:0 Z1:0

.out H
.out i
0 0 -1

H:72 i:105

.include lib
```

prints "Hi".

If all the command definitions described in this paper were defined as macros, the resulting code for any program even a simple one would be enormous. This is because macros are heavily defined through other macros. It means that any command is expanded or inlined at every place where it is used. This is done though all the hierarchy of macro definitions (see Appendix A). To deal with this problem a command can be defined as the actual code working with its own arguments. Such pieces of code are called functions. The macro definition copies the formal arguments to the function's arguments and passes the process control to the function entry point. The caller code also has to pass its current address to enable the process control to be returned back to the caller code. Once the control is returned back from the function, the macro definition can copy the result back to the arguments if necessary. Obviously these functions cannot be recursive because there is no concept of stack[2].

For example, the subtraction `sub` macro and function are defined:

```
.def sub X Y Z : sub_f_X sub_f_Y sub_f_RET sub_f
        .copy X sub_f_X
        .copy Y sub_f_Y
        .copy L sub_f_RET
        0 0 sub_f
        L:J 0
        J:.copy sub_f_X Z
.end

:sub_f: .sub_f_def sub_f_X  sub_f_Y
sub_f_RET:0 sub_f_X:0 sub_f_Y:0

# sub internal macro definition
.def sub_f_def X Y : sub_f_RET

        .copy sub_f_RET Return

        .inv Y
        .inc Y
        .add X Y X

        End:0 0 Return:0

.end
```

First, there is a macro definition, which copies two arguments into global arguments for the function. Next is the global definition of the entry point for the function. The body of the function is defined again through the macro just to keep the internal names outside of the global scope.

---

[2] It does not mean that this concept cannot be introduced. This has been implemented for Higher Subleq.

Ignore for now that a colon precedes the label for the function entry point. The next line defines memory cells for the return value and the two arguments. Two are enough, because the result is passed back inside the first function argument. The next line is a comment. Then there is the body of the function. Its first command is to copy the return address to its last instruction – unconditional jump back to the caller's code[3].

Functions allow the same code to be executed multiple times instead of replicating code in every place where an operation is required. However there is a side effect: since the entry point is global (not inside the macro definition) the code for the function will be present in the program even if this function is not used. This is undesirable. Small programs have to remain small after assembling, and should not include the whole library. To cope with this situation an additional mechanism has been added to the assembler. It marks a command: an instruction or a macro command as conditional if the line begins with a colon. If its name – the label – becomes an unresolved symbol, the command is added to the program. So this is why the line (`sub_f`), in the example above, begins with the colon.

## *Conclusion*

In this paper two goals have been achieved. One is that another OISC language has been invented that seems to have a much simpler instruction than the currently known OISC languages[4]; because it does not explicitly require logic gates.

The other goal has been to prove that bit copying operations coupled with referencing (or addressing) is enough to build a model allowing Turing-complete calculations[5]. It turns out that it is not only possible in principle, but also practically achievable. Simple programs written in this bit-to-bit copying language, work within reasonable time-space resource limits. For example, using the emulator on my PC a program can calculate the factorial of 12 within seconds. The program multiplies numbers from 1 to 12, and then uses modular division to print digits of the result.

The language presented in this paper has been implemented. Its assembler, emulator, and the library can be downloaded from [7].

---

[3] This copy command can be saved if the outer macro can copy directly to this memory cell.
[4] In February 2010 Marc Scibetta published on his web page a model incorporating bit-inversion and a conditional jump.
[5] Only assembly language with a few library macro commands can be regarded exactly as Turing-complete, because they do not have the memory cell size boundary, which limits the address space. Bit copying instructions are loosely Turing-complete or more precisely they are of Linearly Bounded Automaton computational class, which is the class the real computers belong to. Formal proof can be found in [7] where an interpreter of a Turing-complete language DBFI described in [6] is presented. Keymaker (esolangs.org user) argued that the instruction language could be made Turing-complete if addressing is relative, not absolute. It seems that it is possible to redefine the language to use relative addressing, but that is outside of the scope of this paper.

## *Appendix A*

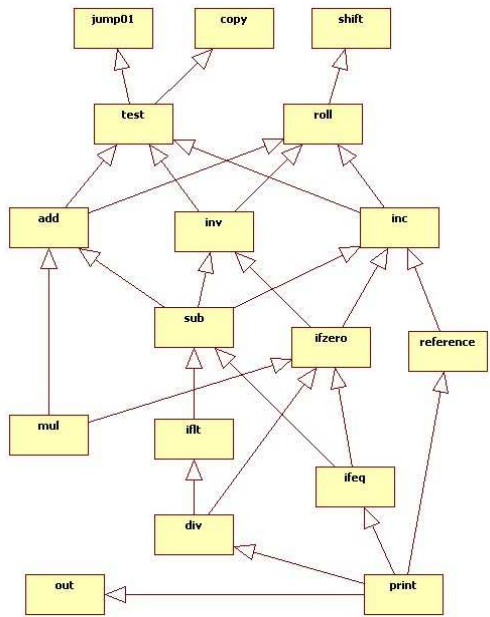

This diagram represents dependencies between functions and macros in the library in the current implementation [7]. Direct dependencies, which are also indirect, are omitted. Different implementation algorithms would result in different dependency diagrams, but general dependency levels would be the same.

## *Appendix B*

### B1 .add

This code defines the addition operation as described in the section "Arithmetic":

```
        .copy ONE ctr
        .copy ZERO adr

begin:  .copy ZERO btr
        .testL adr testx inctestx

inctestx: .inc btr
testx:  .testL X testy inctesty

inctesty: .inc btr
testy:  .testL Y testz inctestz

inctestz: .inc btr
testz:  btr Z
        btr'1 adr'1

        .testH ctr rollcont rollback

rollcont: .shiftL ctr
        .rollR adr
        .rollR X
        .rollR Y
        .rollR Z
        0 0 begin
```

```
rollback: .testL ctr roll End

roll:    .shiftR ctr
         .rollL Z
         0 0 rollback

         End:0 0
         ...
         ctr:0 adr:0 btr:0
```

The ancillary variable `ctr` is used to count the number of rolls applied to the arguments. The variable `adr` is the adder, which is used for passing over bits to the next bit position. The variable `btr` is the sum of three bits taken from the same bit position of the two summing arguments and the adder.

## B2 .inv

The code inverting bits in one word is straightforward. `ctr` is as usual an ancillary variable.

```
         .copy ONE ctr

begin:   .testL ARG copy1 copy0

copy1:   ONE ARG 4?
copy0:   ZERO ARG

         .testH ctr rollcont rollback

rollcont: .shiftL ctr
         .rollR ARG
         0 0 begin

rollback: .testL ctr roll End

roll:    .shiftR ctr
         .rollL ARG
         0 0 rollback

         End:0 0
```

## B3 .div

Below is the working code implementing the division algorithm described in the section "More Arithmetic". Its arguments are: **X** – dividend, **Y** – divisor, **Z** – result of integer division, **R** – remainder.

```
         .copy ZERO Z

         .testH X L1 End
L1:      .testH Y L2 End
L2:      .ifzero Y End begin

begin:   .iflt X Y L3 L4

L3:      .copy X R
         0 0 End

L4:      .copy Y b1
         .copy ONE i1

next:    .copy b1 bp
         .copy i1 ip
         .shiftL b1
```

```
        .shiftL i1

        .iflt X b1 rec L5

rec:    .sub X bp X
        .add Z ip Z
        0 0 begin

L5:     .testH b1 next End

        End:0 0
        ...
        b1:0 bp:0 0
        i1:0 ip:0 0
```

## *Acknowledgements*

I would like to thank my daughter, Sophia Mazonka, for helping me with English grammar. I also would like to thank James Tebneff for valuable comments, which greatly improved the clarity of this paper.